\documentclass[a4paper, 12pt]{article}
  \usepackage{graphicx}
\usepackage{lineno}
\usepackage{hyperref}
\usepackage{amsmath}
 \usepackage[useregional]{datetime2}
 \usepackage{geometry}
\usepackage[english]{babel}

\newenvironment{abstractenv}
  {\begin{center}
   \begin{minipage}{0.85\textwidth}
   \small}
  {\end{minipage}
   \end{center}}
   
 \newcommand{\ignore}[1]{}

\begin{document}
\title{Spin-spin correlations in $\Lambda\bar{\Lambda}$ pair production at the SPD experiment at NICA}
\author{D. Gubachev\textsuperscript{1}, A. Guskov\textsuperscript{1,*}}
\date{\today, \DTMcurrenttime}
\maketitle
\noindent\textsuperscript{1}Joint Institute for Nuclear Research, Dubna, Russia\\
\textsuperscript{*}E-mail: \href{mailto:avg@jinr.int}{avg@jinr.int}

{\footnotesize
\begin{abstractenv}
\linenumbers
\abstract{%
Recent measurements by the STAR collaboration have demonstrated that
spin correlations of $\Lambda\bar{\Lambda}$ pairs provide sensitivity to
the spin structure of $s\bar{s}$ pairs and to spin transfer during hadronization.
 These results, together with recent
CMS measurements at higher energies, open a new avenue for studying spin
dynamics in the transition from partons to hadrons.

A feasibility study of spin-spin correlations in $\Lambda\bar{\Lambda}$
pairs at the SPD experiment at NICA is presented for $pp$ and $dd$
collisions at $\sqrt{s_{NN}}\leq27$~GeV. The expected correlation
$P_{\Lambda\bar{\Lambda}}$ is estimated within the SU(6) and
Burkardt--Jaffe models for a spin-triplet $s\bar{s}$ pair. The kinematic
properties of the process are analysed, and the suitability of the
detector for such measurements is evaluated using a full simulation.
The SPD experiment provides an opportunity to investigate the relative
contributions of non-perturbative and perturbative $s\bar{s}$-production
mechanisms and, with polarized beams, to study the role of initial-state
polarization in spin-dependent partonic dynamics.
}
\end{abstractenv}
}

\section{Introduction}

The study of spin-spin correlations is a powerful tool in modern
physics, providing insight into the nature of interactions across a
wide range of energy scales and offering a unique window into quantum phenomena such as entanglement and decoherence. 
From atomic physics to high-energy
collisions, measurements of spin-dependent observables allow one to
probe the underlying dynamics of a system, test fundamental symmetries.
In high-energy physics, spin correlations between final-state
particles reflect the spin structure of the underlying interactions
and provide a sensitive probe of production and decay dynamics.

The study of spin-spin correlations at the quark level is one of the
actively developing directions of quantum chromodynamics. In 2023,
the ATLAS experiment at the LHC observed for the first time quantum
entanglement in the spin correlations of top--antitop quark pairs
\cite{ATLAS:2023fsd}. Top--antitop pairs in proton--proton collisions
are produced predominantly via gluon fusion ($gg\to t\bar{t}$).
Near the production threshold, the spin density matrix of the
$t\bar{t}$ system is dominated by a spin-singlet component, which
corresponds to a maximally entangled state. This was the first
observation of quantum entanglement for a quark pair.
These results have been confirmed and extended by measurements at CMS
\cite{CMS:2024pts,Yazgan:2025pah}.

The analysis of spin correlations at the level of secondary hadrons takes research to a new level. Thus, in the BESIII experiment, in the cascade decay $J/\psi \to \gamma \eta_c$, $\eta_c \to \Lambda \bar{\Lambda}$, a violation of Bell-like inequalities was observed for the first time in high-energy physics with a significance of more than $5\sigma$ \cite{BESIII:2025vsr}. This result
provides a stringent test of local realism and constitutes evidence
for quantum entanglement in the final-state hyperon pair. 

The STAR experiment subsequently observed significant spin-spin
correlations between $\Lambda$ and $\bar{\Lambda}$ hyperons in
proton--proton collisions, consistent with the interpretation that
spin correlations of $s\bar{s}$ pairs produced from the QCD vacuum in
a spin-triplet state can survive hadronization and be transferred to
the final-state hyperons \cite{STAR:2025njp}. The correlation was found
to be larger for smaller angular separations between the hyperons.
Thus, measurements of spin correlations in hadronic final states
provide a new way to investigate the interplay between quantum
correlations and non-perturbative QCD dynamics during hadronization.

\section{Spin correlations in strangeness production in hadronic collisions}

 \begin{figure}[!b]
  \begin{center}
    \includegraphics[width=1.\textwidth]{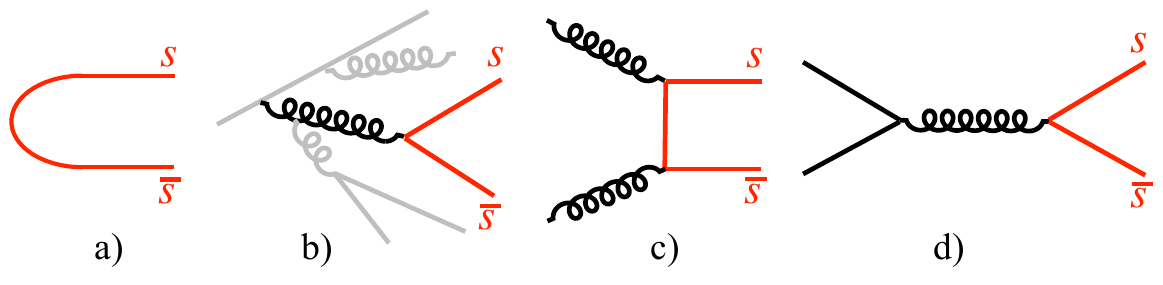}
  \end{center}
  \caption{$s\bar{s}$-pair production mechanisms: materialisation of a triplet spin state from the QCD vacuum condensate (a);  gluon splitting in the parton shower evolution (b); production in a hard interaction: gluon-gluon fusion (c) and quark-antiquark annihilation (d).}
  \label{diagrams}
\end{figure}

In proton--proton collisions at the collision energy studied by STAR
($\sqrt{s}=200$~GeV) and at lower energies, several mechanisms of
$s\bar{s}$ pair production can be relevant.
In the non-perturbative regime, $s\bar{s}$ pairs originating from the
QCD vacuum (Fig. \ref{diagrams}(a)) are expected to inherit the vacuum quantum numbers,
$J^{PC}=0^{++}$, which correspond to a ${}^3P_0$ state for a
quark--antiquark pair. Another mechanism is
gluon splitting in the parton shower, $g\to s\bar{s}$ (Fig. \ref{diagrams}(b)), which can
lead to an opposite-sign spin correlation. Both mechanisms
can produce $s\bar{s}$ pairs with small angular separation. In
addition, hard processes such as gluon--gluon fusion,
$gg\to s\bar{s}$, and quark--antiquark annihilation,
$q\bar{q}\to s\bar{s}$ (Fig. \ref{diagrams}(c) and (d), respectively), may contribute to the observed spin-spin
correlations and can populate larger angular separations.

The possibility of using $\Lambda\bar{\Lambda}$ spin correlations to probe the
mechanism of $s\bar{s}$ pair production at the quark level in hadronic
collisions was investigated in Ref.~\cite{Ellis:2011kq}. In that work,
several perturbative production mechanisms were considered, including
gluon splitting $g\to s\bar{s}$ and gluon--gluon fusion
$gg\to s\bar{s}$. The resulting spin correlation depends on the spin state of the
produced $s\bar{s}$ pair: a ${}^3S_1$ state gives a positive
correlation, whereas a ${}^1S_0$ state gives a negative correlation.
 These different spin-correlation patterns provide
a basis for studying the underlying mechanisms of $s\bar{s}$ pair
production experimentally.

More recently, it was proposed in Ref.~\cite{Gong:2021bcp} that
$\Lambda\bar{\Lambda}$ spin correlations can be used to probe quantum
entanglement and to test Bell-type inequalities at high-energy colliders,
establishing a direct connection between QCD dynamics and quantum
information. The study also showed that the degree of quantum
entanglement decreases with increasing hadron multiplicity, while the
spin correlation itself can remain sizeable. Since the average hadron
multiplicity generally increases with collision energy, the lower-energy
regime accessible at NICA provides a complementary environment for
studying quantum correlations with reduced multiplicity-induced
dilution.

Measurement of spin-spin correlations of hyperon pairs in the final state allows one to probe both the production mechanism of $s\bar{s}$ pairs and their spin state, as well as the transfer of spin correlations from quarks to hyperons during hadronization. 
The $\Lambda\bar{\Lambda}$ pair is particularly suitable for such a
probe because the $\Lambda$ hyperon has a self-analyzing weak decay
$\Lambda\to p\pi^-$ with a decay asymmetry parameter
$\alpha_-\approx0.75$ and a branching fraction of 63.9\%.
However, the situation is complicated by the fact that a significant fraction of strange quarks hadronize into strange hadrons heavier than the $\Lambda$, which subsequently decay into $\Lambda$ hyperons (feed-down contribution). In the STAR kinematics, the fraction of $\Lambda\bar{\Lambda}$ pairs in which both hyperons are primary (i.e., produced directly in the fragmentation process rather than through decays of heavier strange hadrons) is only about 11\%.

For reconstructed decays $\Lambda \to p \pi^-$ and $\bar{\Lambda} \to \bar{p} \pi^+$, the (anti-)protons were boosted into the rest frames of their parent hyperons, and the angle $\theta^\star$ between the two boosted (anti-)protons was determined. The angular distribution was expected to be the following:
\begin{equation}
\frac{1}{N} \frac{dN}{d \cos\theta^{\star}} = \frac{1}{2}(1+\alpha_- \alpha_+ P_{\Lambda\bar{\Lambda}} \cos \theta^{\star}).
\end{equation}
Here $P_{\Lambda\bar{\Lambda}}$ is the spin correlation signal, related with the correlation  matrix $C_{ij}$ as
\begin{equation}
P_{\Lambda\bar{\Lambda}}=\frac{1}{3}Tr(C_{ij})=\frac{1}{3}(C_{xx}+C_{yy}+C_{zz}).
\end{equation}
The parameters $\alpha_-$ and $\alpha_+$ are $\Lambda$ and $\bar{\Lambda}$ decay parameters ($\alpha_-$ = 0.747 $\pm$ 0.009, $\alpha_+$ = -0.757$\pm$0.004). 
For a pure spin-triplet $s\bar{s}$ state, the spin-correlation signal averaged over the three spatial directions is $P_{\Lambda\bar{\Lambda}}=1/3$, whereas for a spin-singlet state it is $P_{\Lambda\bar{\Lambda}}=-1$.  
Similar analysis was also performed for sets of $\Lambda\Lambda$ and $\bar{\Lambda}\bar{\Lambda}$ pairs as well as $K^0_s K^0_s$ pairs to control systematic effects. This control sample is insensitive to spin correlations, as the $K^0_s$ is a spin-0 meson, and serves to validate the background estimation procedure.

The SU(6) quark model \cite{SU6} and the Burkardt–Jaffe phenomenological model \cite{BJ} were used to describe the transfer of spin polarization from quarks to final-state $\Lambda$ and $\bar{\Lambda}$ hyperons including their production through higher states. The maximum expected values of the  spin-spin correlation $P_{\Lambda\bar{\Lambda}}$ were estimated to be $0.096\pm0.004$ and $0.015\pm0.002$ for the SU(6) and the Burkardt–Jaffe models, respectively.

The STAR experiment observed for the first time a significant spin-spin
correlation between $\Lambda$ and $\bar{\Lambda}$ hyperons:
$P_{\Lambda\bar{\Lambda}}=0.181\pm0.035_{\mathrm{stat}}
\pm0.022_{\mathrm{syst}}$. This is consistent with the survival and transfer of spin correlations from a triplet $s\bar{s}$ state produced non-perturbatively from the QCD vacuum, although it does not by itself establish quantum entanglement of the initial $s\bar{s}$ pair.
 A key observation is that the
correlation is observed for hyperons emitted at small angular
separations, $|\Delta y|<0.5$ and $|\Delta\phi|<\pi/3$, and decreases
with increasing angular separation. This result provides a new
experimental tool to probe non-perturbative QCD, enabling the study of
confinement dynamics and the role of spin degrees of freedom in the
transition from quarks to hadrons.

Recently, the CMS collaboration at the LHC has significantly extended
these studies by reporting the first measurement of spin-spin
correlations of hyperon pairs in $pp$ collisions at $\sqrt{s}=13$~TeV
and in $p\mathrm{Pb}$ collisions at $\sqrt{s_{NN}}=8.16$~TeV
\cite{CMS:2026edg}. In contrast to the positive correlation observed by
STAR for $\Lambda\bar{\Lambda}$ pairs at $\sqrt{s}=200$~GeV, CMS found
negative values for $\Lambda\Lambda$ and $\bar{\Lambda}\bar{\Lambda}$
pairs at small angular separations, of about $-0.1$, while no
significant signal was observed for $\Lambda\bar{\Lambda}$ pairs.

An attempt to explain the observed dependence of the spin
correlation on the angular separation $\Delta R = \sqrt{\Delta y^2 + \Delta \phi ^2}$ was made in
Ref.~\cite{Barata:2026hck} within the framework of open quantum systems.
In this phenomenological approach, the $s\bar{s}$ pair is treated as an
open quantum system interacting with the surrounding hadronic
environment during hadronization. The loss of spin information is
described by a depolarizing channel with a separation-dependent
decoherence rate, leading to an exponential suppression of
$P_{\Lambda\bar{\Lambda}}$ with increasing $\Delta R$. The model provides
a reasonable description of both the STAR and CMS data. Within this
framework, the observed decrease of the spin correlation with
$\Delta R$ is interpreted as a manifestation of quantum decoherence
during the transition from quarks to hadrons.

These results suggest that the mechanisms governing hyperon
production and spin transfer may depend on the collision energy and
collision system, making the intermediate-energy range accessible at
NICA particularly relevant for further studies.

\section{Physics prospects for spin-spin correlations in  $\Lambda\bar{\Lambda}$ pairs production at SPD}

The Spin Physics Detector (SPD) project at the NICA collider plans to
study the spin structure of nucleons and spin-dependent effects in
collisions of polarized protons and deuterons at energies up to
27~GeV \cite{Arbuzov:2020cqg,Abramov:2021vtu}. Already in the first
phase of the experiment, the SPD tracking system will be available,
consisting of the main tracker based on straw tubes and the central
tracker based on Micromegas detectors, placed in the magnetic field of
a superconducting solenoid with a field of 1~T \cite{SPD:2024gkq}.
Such a configuration of the experimental setup will make it possible
to study spin-spin correlations in hyperon production from the
beginning of the experiment. In its complete configuration, the detector will feature improved tracking performance and enhanced capabilities for secondary particle identification.

The study of $\Lambda\bar{\Lambda}$ spin correlations at SPD can address
several complementary questions. First, the lower collision energies
and, consequently, lower event multiplicities accessible at NICA provide
a complementary environment for searching for quantum entanglement in
the $\Lambda\bar{\Lambda}$ system and for studying its relation to the
spin density matrix. Second, measurements of the spin correlation as a
function of the angular separation $\Delta R$ at energies below those
explored by STAR and CMS will provide a new test of the mechanisms
responsible for spin decoherence during hadronization. The dependence
on event multiplicity can further help to disentangle decoherence effects
associated with the hadronic environment. Third, the energy, kinematic,
and beam-polarization dependences of the correlations can provide
sensitivity to different mechanisms of $s\bar{s}$ production, including
non-perturbative production, gluon splitting, and hard scattering.
Finally, measurements with different $\Lambda$ parentage and
spin-dependent observables can provide information on the transfer of
the $s$-quark polarization to the final-state hyperons. 

The possibility of using beam polarization is particularly important for
the latter two questions. In collisions of longitudinally polarized protons or deuterons, the
spin-spin correlations of $\Lambda\bar{\Lambda}$ pairs may acquire a
dependence on the beam spin configuration. Different mechanisms of
$s\bar{s}$-pair production can respond differently to the polarization
of the initial partons. The production of $s\bar{s}$ pairs through a
non-perturbative mechanism associated with the QCD vacuum is not
expected to have a direct dependence on the beam polarization, whereas
spin-dependent hard processes and the spin-dependent partonic structure of
the initial state can introduce such a dependence. Spin-dependent
effects in the parton shower may further modify the observed
correlations. Thus, measurements with longitudinally polarized beams
at SPD provide an opportunity not only to study hyperon spin-spin
correlations, but also to investigate the role of initial-state
polarization and spin-dependent partonic dynamics in their production.
In transversely polarized collisions, spin-dependent hard processes
and parton-shower dynamics can give rise to azimuthal modulations of
the measured $\Lambda\bar{\Lambda}$ spin correlations.

\section{Expectations and detector performance}

\begin{figure}[!h]
     \begin{minipage}[ht]{0.45\linewidth}
       \center{\includegraphics[width=1\textwidth]{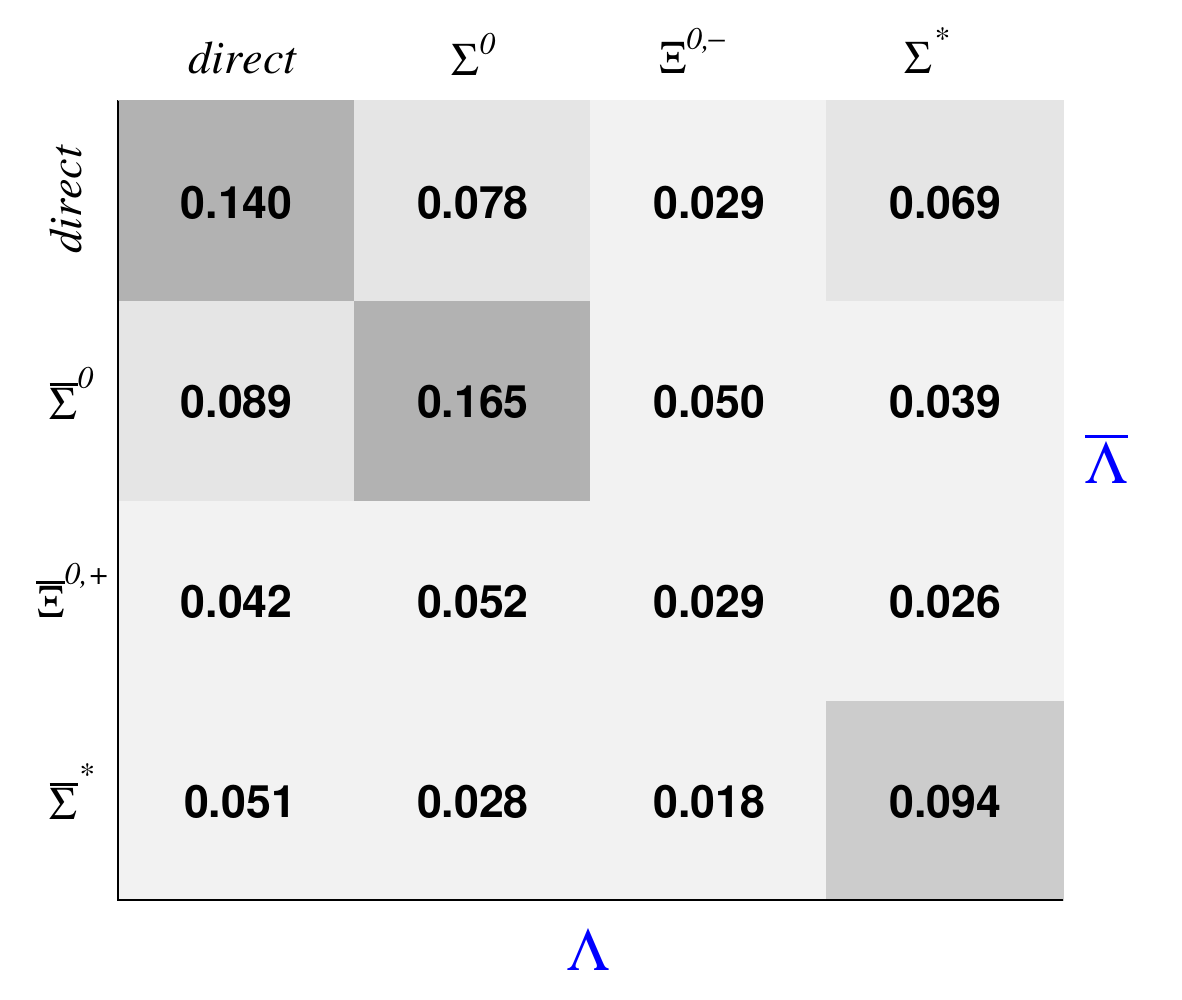} \\ (a)}
  \end{minipage}
    \hfill
   \begin{minipage}[ht]{0.55\linewidth}
    \center{\includegraphics[width=1\textwidth]{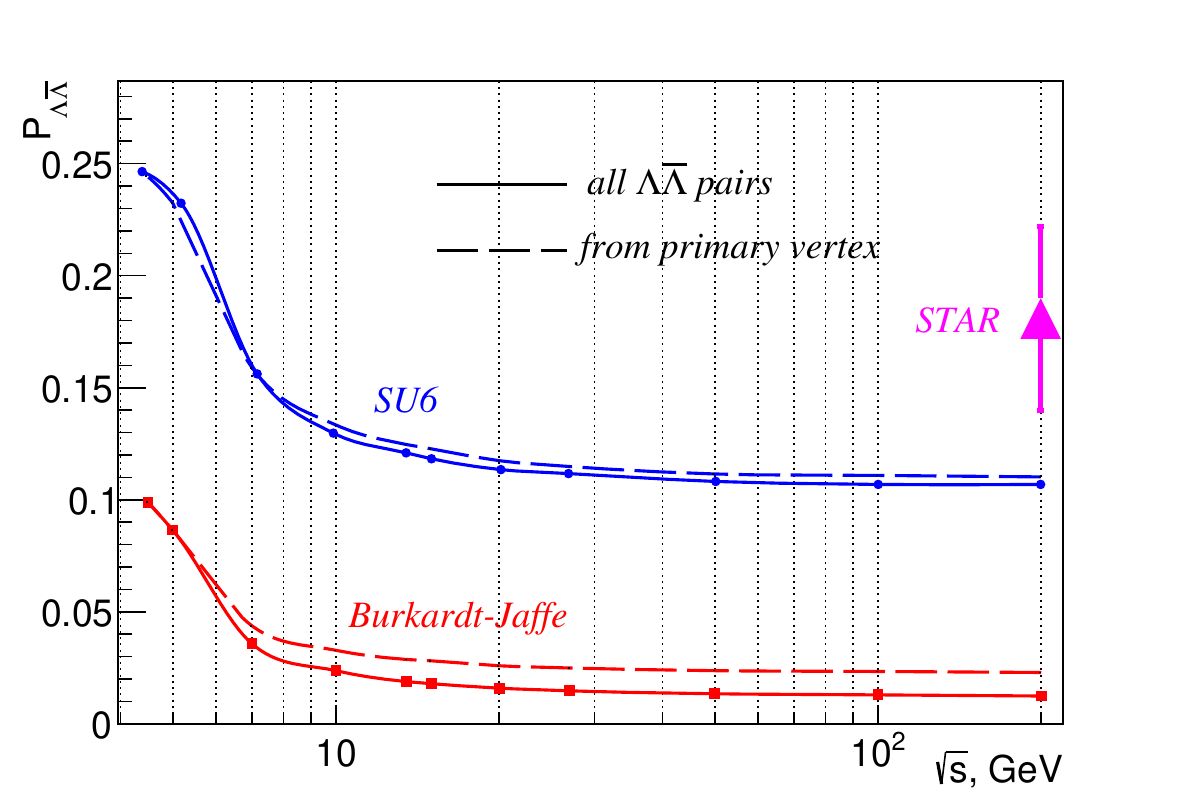} \\ (b)}
  \end{minipage}
  \caption{Relative contributions $R_{ij}$ of different sources of $\Lambda\bar{\Lambda}$ pairs in $pp$ collisions at $\sqrt{s}=$27 GeV (a). Maximum correlation $P_{\Lambda\bar{\Lambda}}$ as a function of collision energy $\sqrt{s}$ estimated within the SU(6) and BJ models for all pairs (solid) and for the case when at least one hyperon does not point to the primary vertex (dashed) (b).}
  \label{corr}
\end{figure}

The maximum expected $\Lambda\bar{\Lambda}$ spin-spin correlation for a
spin-triplet $s\bar{s}$ pair, based on the SU(6) quark model and the
Burkardt--Jaffe model, is calculated according to
\begin{equation}
P_{\Lambda\bar{\Lambda}}
=
\frac{1}{3}
\sum_{i,j}
R_{ij}\,
P^i_{\Lambda}
P^j_{\bar{\Lambda}},
\end{equation}
where $R_{ij}$ is the fraction of $\Lambda\bar{\Lambda}$ pairs in which
the $\Lambda$ and $\bar{\Lambda}$ belong to production categories $i$ and $j$, respectively, as estimated using the
\textsc{Pythia} 8.312 event generator \cite{Bierlich:2022pfr}
(see Fig.~\ref{corr}(a)). The factor $1/3$ reflects the isotropic average over the three spin directions for the initial triplet state. The quantities $P^i_{\Lambda}$
($P^j_{\bar{\Lambda}}$) denote the contribution of the $s$
($\bar{s}$)-quark spin to the spin of a $\Lambda$ ($\bar{\Lambda}$)
hyperon produced either directly or through the decay of a heavier
state (see Table~\ref{Tab1}).

Figure \ref{corr}(b) shows the maximum correlation $P_{\Lambda\bar{\Lambda}}$ as a function of proton-proton collision energy $\sqrt{s}$, calculated within the SU(6) and Burkardt-Jaffe (BJ) models. The solid curves represent the total correlation including all pairs. 
The increase in the correlation at lower collision energies is driven
by the rapid decrease of the feed-down contribution. 
The dashed curves correspond to the scenario where pairs in which at least one hyperon does not point to the primary vertex are excluded; in the simulation, these pairs are predominantly associated with $\Xi$  feed-down.
      As one can see, the $\Xi$ feed-down contribution significantly modifies
the prediction of the BJ model due to the negative sign of the
corresponding spin-transfer coefficient $P_{\Lambda(\bar{\Lambda})}$,
while its effect on the SU(6) model prediction is negligible.
Assuming that nuclear effects are negligible, a similar spin-transfer
pattern is expected for $dd$ collisions as a consequence of isospin
symmetry.

\begin{figure}[!h]
     \begin{minipage}[ht]{0.5\linewidth}
       \center{\includegraphics[width=1\textwidth]{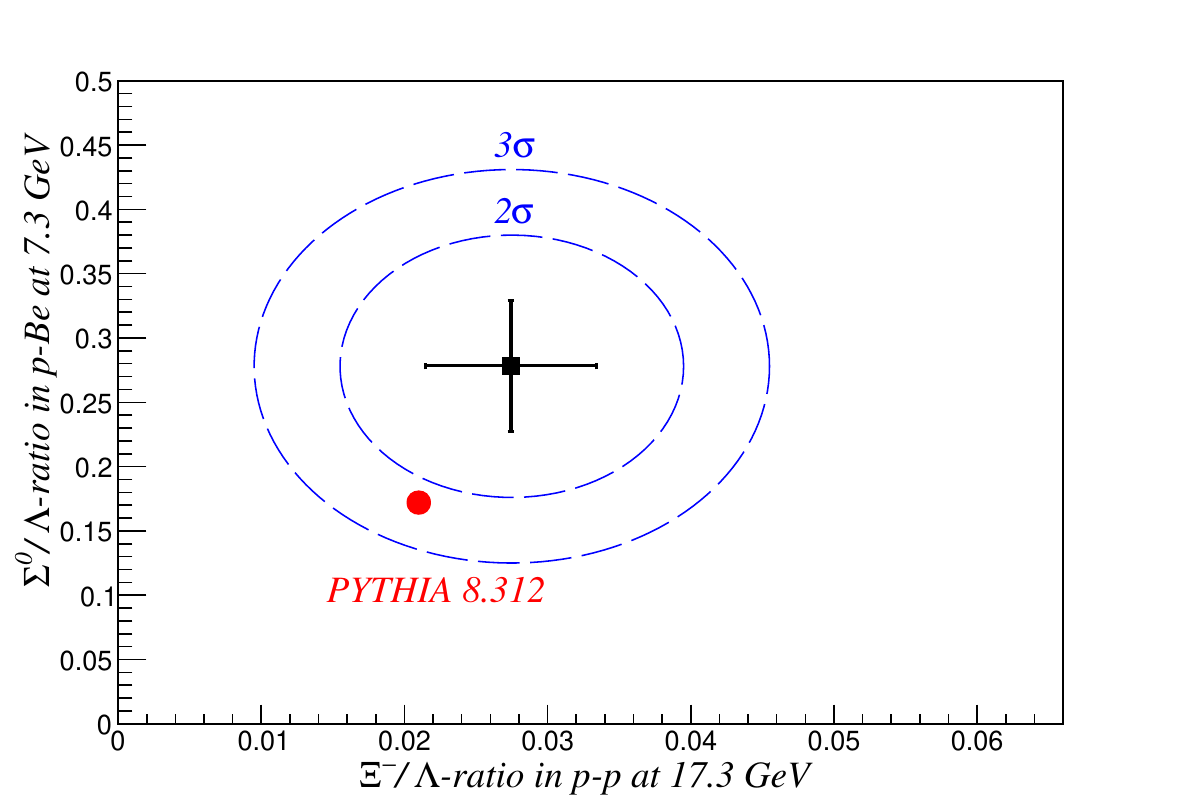} \\ (a)}
  \end{minipage}
    \hfill
   \begin{minipage}[ht]{0.5\linewidth}
    \center{\includegraphics[width=1\textwidth]{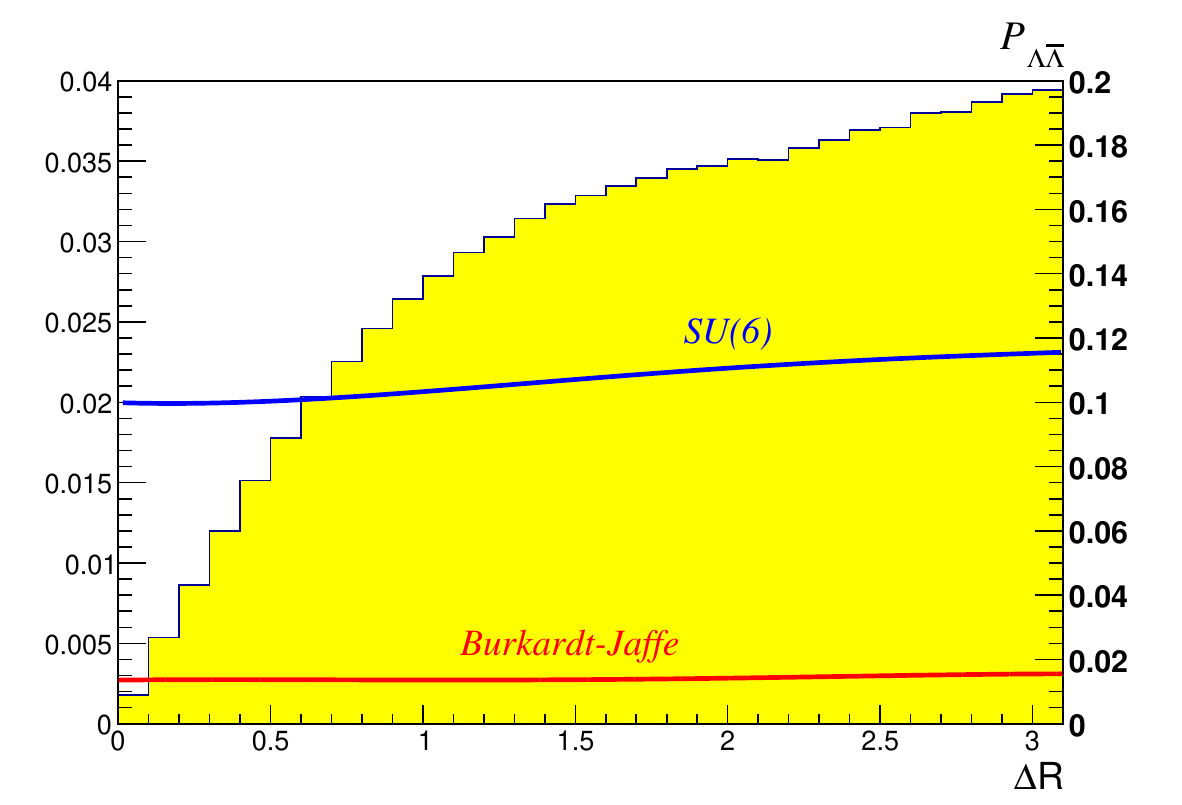} \\ (b)}
  \end{minipage}
  \caption{Experimental measurements for  ratios of hyperon yields with uncertainties are shown together with the corresponding \textsc{Pythia} 8.312 predictions (a). 
  Angular separation $\Delta R$ of $\Lambda$ and $\bar{\Lambda}$  distribution for $\sqrt{s}=27$ GeV (b). }
  \label{validation}
\end{figure}

As a consistency check of the PYTHIA-based description of strange-hadron production, we compare the simulated hyperon yield ratios with available experimental measurements. Figure~\ref{validation}(a) shows the ratios $\Xi^-/\Lambda$ and $\Sigma^0/\Lambda$ measured at $\sqrt{s}=17.3$ and $7.3$~GeV, respectively, including their experimental uncertainties (see also Tab. \ref{Tab2}, together with the corresponding PYTHIA predictions. The agreement is not expected to be exact, in particular given the differences between the collision systems and kinematic conditions of the available measurements and the present simulation. Nevertheless, the comparison indicates that the PYTHIA description provides a reasonable baseline for the relative production of strange hyperons relevant for the present study.

\begin{table}[htp]
\caption{Contribution of the $s$-quark spin to the spin of a $\Lambda$ hyperon produced either directly or through the feed-down mechanism calculated within the SU(6) and Burkardt–Jaffe models  \cite{Ellis:2007ig}}.
\begin{center}
\begin{tabular}{|c|c|c|c|c|}
\hline
$\Lambda$'s parent & $c\tau$, & Main decay& $P_{\Lambda(\bar{\Lambda})}$, SU(6) & $P_{\Lambda(\bar{\Lambda})}$, Burkardt– \\
                               & cm & channel, \% &  model & Jaffe model \\
\hline
$s$-quark & - & - & 1 &  0.63 \\
$\Sigma^0$ & $2.2\times10^{-9}$ & $\Lambda\gamma$, 100 & 1/9 & 0.15 \\
$\Xi^{0}$ & 8.7 & $\Lambda\pi^0$, 99.5 & 0.6 & -0.37 \\
$\Xi^{-}$ & 4.9 & $\Lambda\pi^-$, 99.9&0.6 & -0.37 \\
$\Sigma^*(1385)$ & $7\times10^{-13}$  & $\Lambda\pi$, 87 & 5/9 &  - \\
\hline
\end{tabular}
\end{center}
\label{Tab1}
\end{table}%

\begin{table}[htp]
\caption{Ratios of hyperon yields measured in the experiment as compared with those obtained from \textsc{Pythia}.}
\begin{center}
\begin{tabular}{| l | c | c | c |}
\hline
Ratio &  $\sqrt{s}$, GeV & Measurement & \textsc{Pythia} 8.312 \\
\hline
$\Xi^-/\Lambda$ & 17.3 (p-p)  & 0.028$\pm0.0060$ \cite{NA61SHINE:2015haq,NA61SHINE:2020dwg} & 0.021 \\  
$\Sigma^0/\Lambda$ &7.3 (p-Be) &  0.278$\pm$0.051 \cite{Dukes:1988ir}  & 0.171\\ 
\hline
\end{tabular}
\end{center}
\label{Tab2}
\end{table}%

 \begin{figure}[!h]
     \begin{minipage}[ht]{0.47\linewidth}
       \center{\includegraphics[width=1\textwidth]{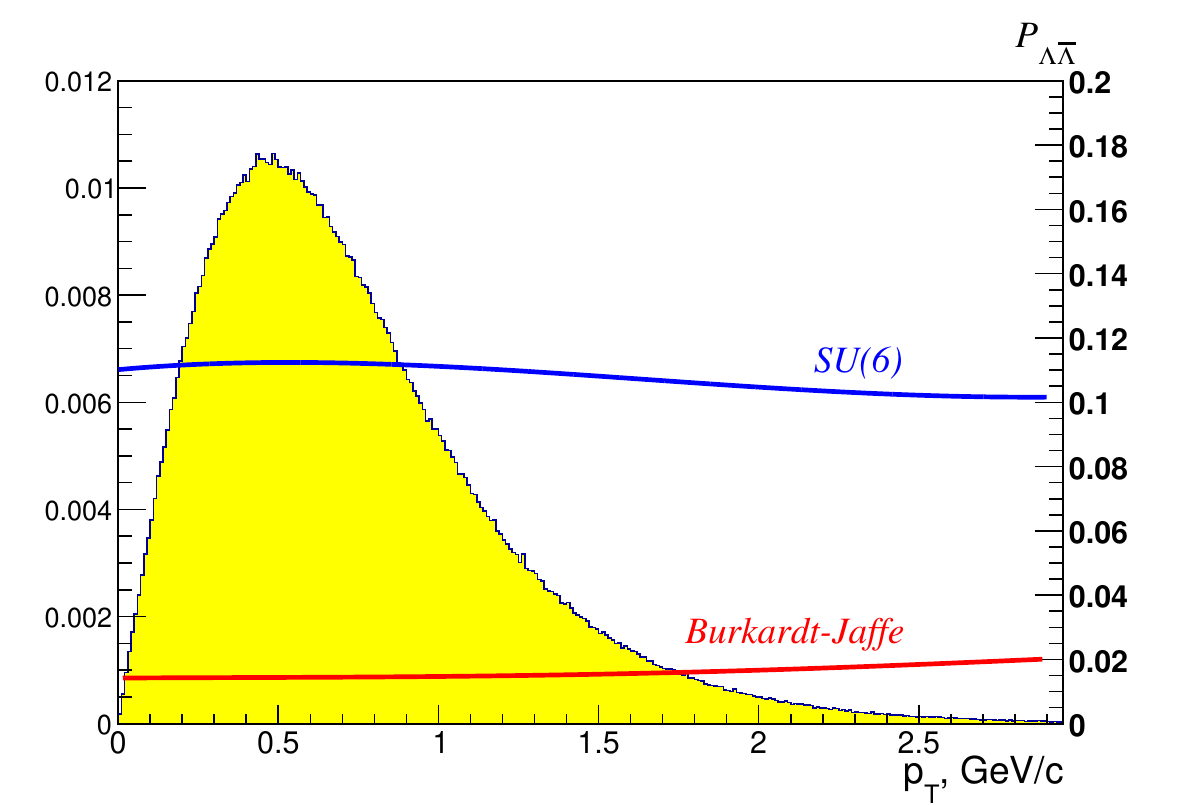} \\ (a)}
  \end{minipage}
    \hfill
   \begin{minipage}[ht]{0.53\linewidth}
    \center{\includegraphics[width=1\textwidth]{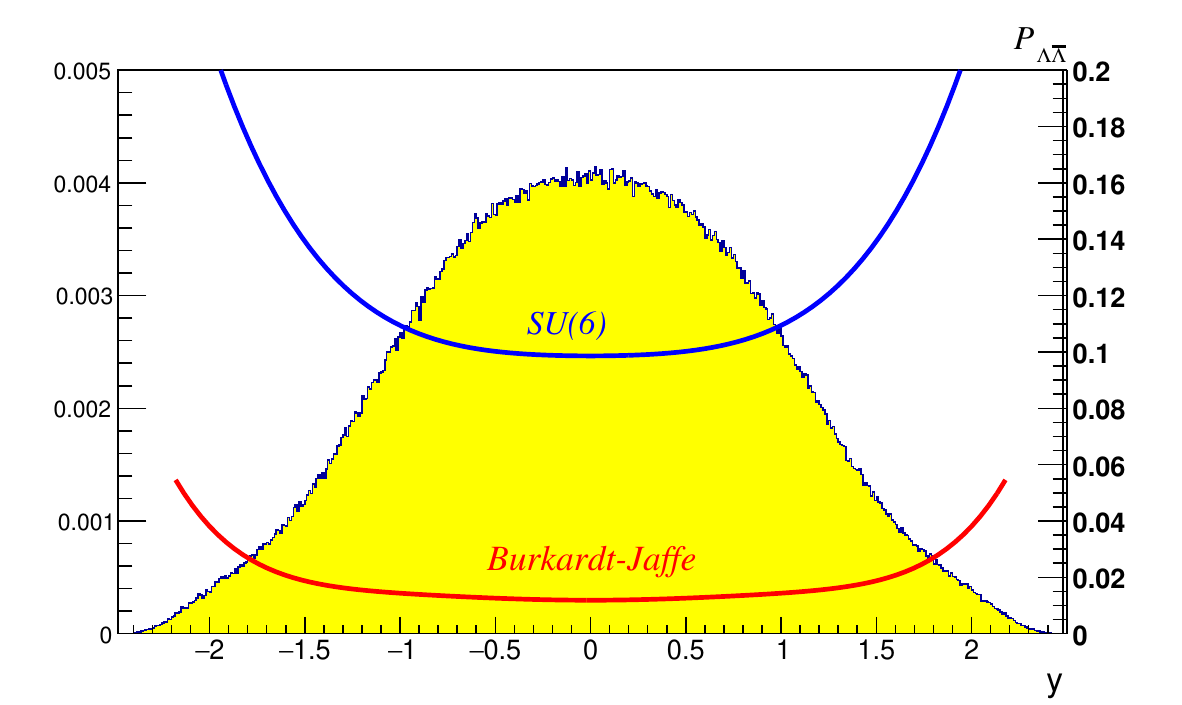} \\ (b)}
  \end{minipage}
  \caption{Distributions for transverse momentum $p_T$ (a) and rapidity $y$ (b) of $\Lambda \bar{\Lambda}$ pairs for $\sqrt{s}=27$ GeV.}
  \label{Y_pt}
\end{figure}


Figure~\ref{validation}(b) shows the angular separation $\Delta R$ of the 
$\Lambda\bar{\Lambda}$ pairs, while Fig.~\ref{Y_pt} presents 
the distributions of the pairs in transverse momentum $p_T$ and rapidity $y$ 
 for $\sqrt{s}=27$~GeV. The right-side axes of these figures also 
show the corresponding maximum spin correlation $P_{\Lambda\bar{\Lambda}}$ 
as a function of these variables, evaluated within the SU(6) and 
BJ models. The notable dependence of $P_{\Lambda\bar{\Lambda}}$ 
on rapidity can be attributed to the fact that pairs with a 
feed-down contribution tend to be emitted, on average, at larger 
angles relative to the beam axis.

Having determined the correlation parameter $P_{\Lambda\bar{\Lambda}}$, we still cannot make a definite conclusion about all elements of the spin correlation matrix $C_{ij}$. In the general polarized case, a complete characterization of the $\Lambda\bar{\Lambda}$ spin state requires reconstruction of the full spin density matrix, including the single-particle polarization vectors and all components of the spin correlation tensor. This would require measuring the corresponding angular correlations in the hyperon weak decays. In the present study, we consider a reduced description based on transverse rotational invariance, which allows the number of independent correlation parameters to be reduced. In this description, single-particle polarization effects are neglected, and only the spin-correlation tensor is retained.
Assuming that the spin correlation tensor is diagonal and invariant under rotations in the transverse plane, the off-diagonal elements vanish and the transverse components are equal, $C_{xx}=C_{yy}=C_T$, while the longitudinal component is denoted by $C_{zz}=C_L$. The correlation parameter is then related to these components by
\begin{equation}
2C_T+C_L=3P_{\Lambda\bar{\Lambda}}.
\end{equation}
In this case, the spin density matrix can be written as:
\begin{equation}
\rho = 
\begin{pmatrix}
\dfrac{1 + C_L}{4} & 0 & 0 & 0 \\[6pt]
0 & \dfrac{1 - C_L}{4} & \dfrac{3P_{\Lambda\bar{\Lambda}} - C_L}{4} & 0 \\[6pt]
0 & \dfrac{3P_{\Lambda\bar{\Lambda}} - C_L}{4} & \dfrac{1 - C_L}{4} & 0 \\[6pt]
0 & 0 & 0 & \dfrac{1 + C_L}{4}
\end{pmatrix},
\end{equation}
The partial transpose of this matrix with respect to the second subsystem yields the following set of eigenvalues, the non-negativity of which is a necessary and sufficient condition \cite{Peres, HHH} for the absence of quantum entanglement of the hyperons:
\[
\begin{aligned}
\lambda_1 &= \frac{1 + 3P_{\Lambda\bar{\Lambda}}}{4}, \\
\lambda_2 &= \frac{1 + 2C_L - 3P_{\Lambda\bar{\Lambda}}}{4}, \\
\lambda_3 &= \lambda_4 = \frac{1 - C_L}{4}.
\end{aligned}
\]

\begin{figure}[!h]
     \begin{minipage}[ht]{0.55\linewidth}
       \center{\includegraphics[width=1\textwidth]{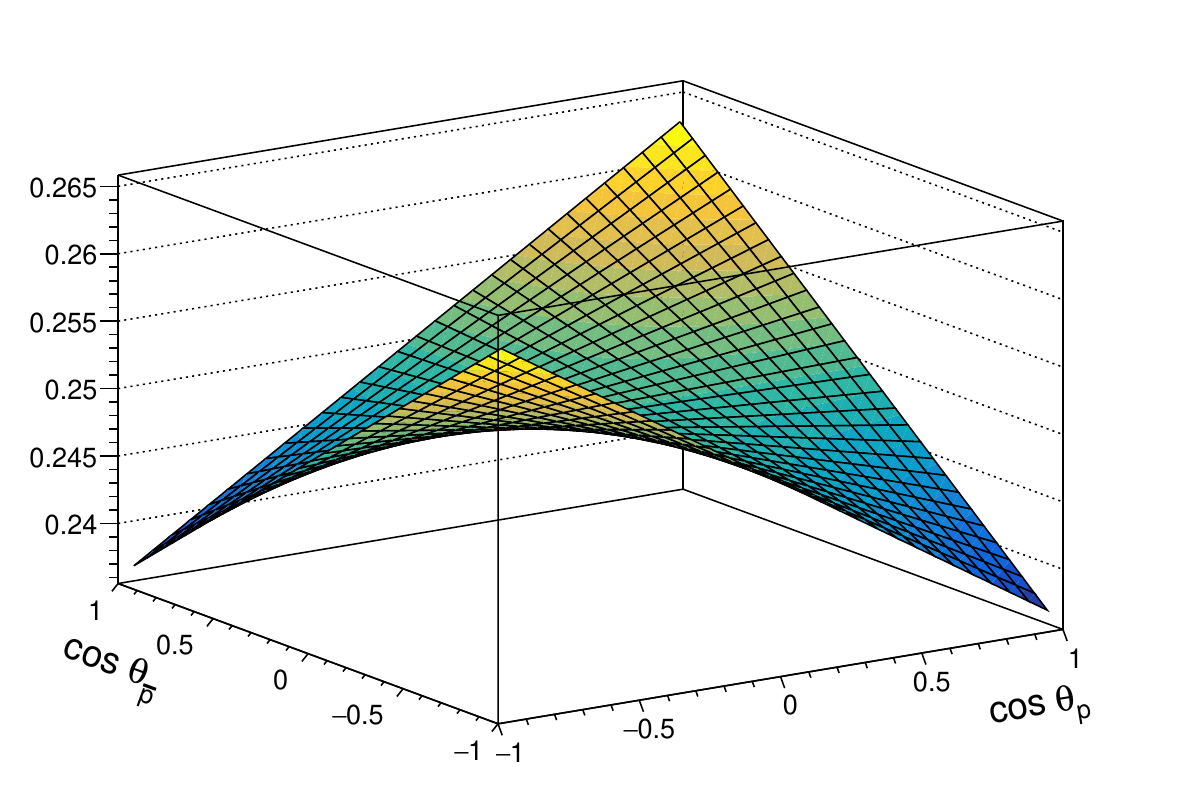} \\ (a)}
  \end{minipage}
    \hfill
   \begin{minipage}[ht]{0.45\linewidth}
    \center{\includegraphics[width=1\textwidth]{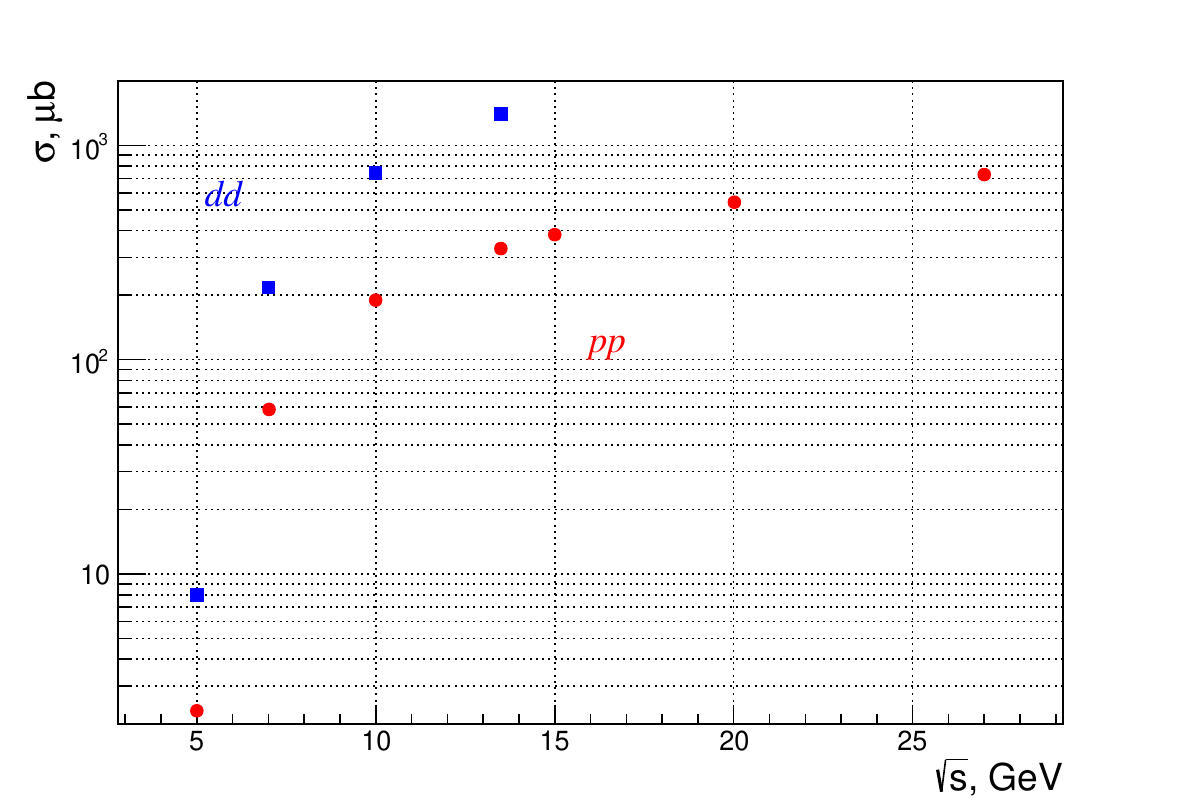} \\ (b)}
  \end{minipage}
  \caption{ Correlation of the emission angles of the proton and antiproton with respect to the $Z$ axis in the rest frames of the $\Lambda$ and $\bar{\Lambda}$, respectively, for $C_L = 0.1$ (a).
  Inclusive cross section for $\Lambda\bar{\Lambda}$ pair production as a function of  collision energy $\sqrt{s}_{NN}$ for $pp$ and $dd$ collisions (b). }
  \label{section}
\end{figure}

For $P_{\Lambda\bar{\Lambda}} = 0.1$, entanglement of the final-state $\Lambda\) and \(\bar{\Lambda}$ occurs only for $C_L < -0.35$. The longitudinal spin-spin correlation $C_{L}$ is experimentally accessible through the mean value of the product of the proton and antiproton angular distributions:
\begin{equation}
\frac{d^2N}{d(\cos\theta_p)\,d(\cos\theta_{\bar p})}
=
\frac{1}{4}
\left[
1+\alpha_\Lambda\alpha_{\bar\Lambda}
C_{L}\,
\cos\theta_p\cos\theta_{\bar p}
\right],
\end{equation}
where $\theta_p$ ($\theta_{\bar{p}}$) is the angle between the proton (antiproton) momentum and the $z$-axis in the corresponding $\Lambda$ ($\bar{\Lambda}$) rest frame. The corresponding distribution for $C_L=0.1$ is shown in Fig. \ref{section}(a). So, 
\begin{equation}
C_{L} = \frac{9}{\alpha_{\Lambda}\alpha_{\bar{\Lambda}}}
\left\langle \cos\theta_p \cos\theta_{\bar{p}} \right\rangle.
\end{equation}

Under the assumption of transverse rotational invariance, $C_{xx}=C_{yy}=C_T$. Therefore, if the measured value satisfies $C_{L}\approx P_{\Lambda\bar{\Lambda}}$, it follows from
\begin{equation}
2C_T+C_{L}=3P_{\Lambda\bar{\Lambda}}
\end{equation}
that $C_T\approx C_{L}\approx P_{\Lambda\bar{\Lambda}}$, corresponding to an isotropic spin-correlation tensor. For $P_{\Lambda\bar{\Lambda}}>-1/3$, a negativity $\mathcal{N}\equiv \sum_{\lambda_i<0}|\lambda_i|$ can be used as a  quantitative measure of entanglement, which in this case is given by:
\begin{equation}
\mathcal{N} = \max\left[
0,
\frac{3P_{\Lambda\bar{\Lambda}}-1-2C_L}{4}
\right].
\end{equation}

 Such an isotropic spin-correlation structure is expected at the $s\bar{s}$ level for the vacuum ${}^3P_0$ production mechanism. Thus, a significant deviation from the relation $P_{\Lambda\bar{\Lambda}}=C_L$
would indicate a departure from this isotropic spin-correlation structure in the final state. Such a deviation could result either from contributions of other $s\bar{s}$ production mechanisms or from spin-dependent effects arising during hadronization. 

The expected inclusive cross section for $\Lambda\bar{\Lambda}$ pair production as a function of  collision energy for $pp$ and $dd$ collisions, according to \textsc{Pythia}, is shown in Fig. \ref{section}(b). For the estimation of the cross section for $dd$ collisions, the Angantyr framework within \textsc{Pythia} was employed. This framework uses a Glauber-based initial-state description to determine the number and geometry of nucleon–nucleon subcollisions and does not take into account collective behaviour of nucleons that may lead to a slight underestimation of the cross section near the threshold.

 \begin{figure}[!h]
     \begin{minipage}[ht]{0.5\linewidth}
       \center{\includegraphics[width=1\textwidth]{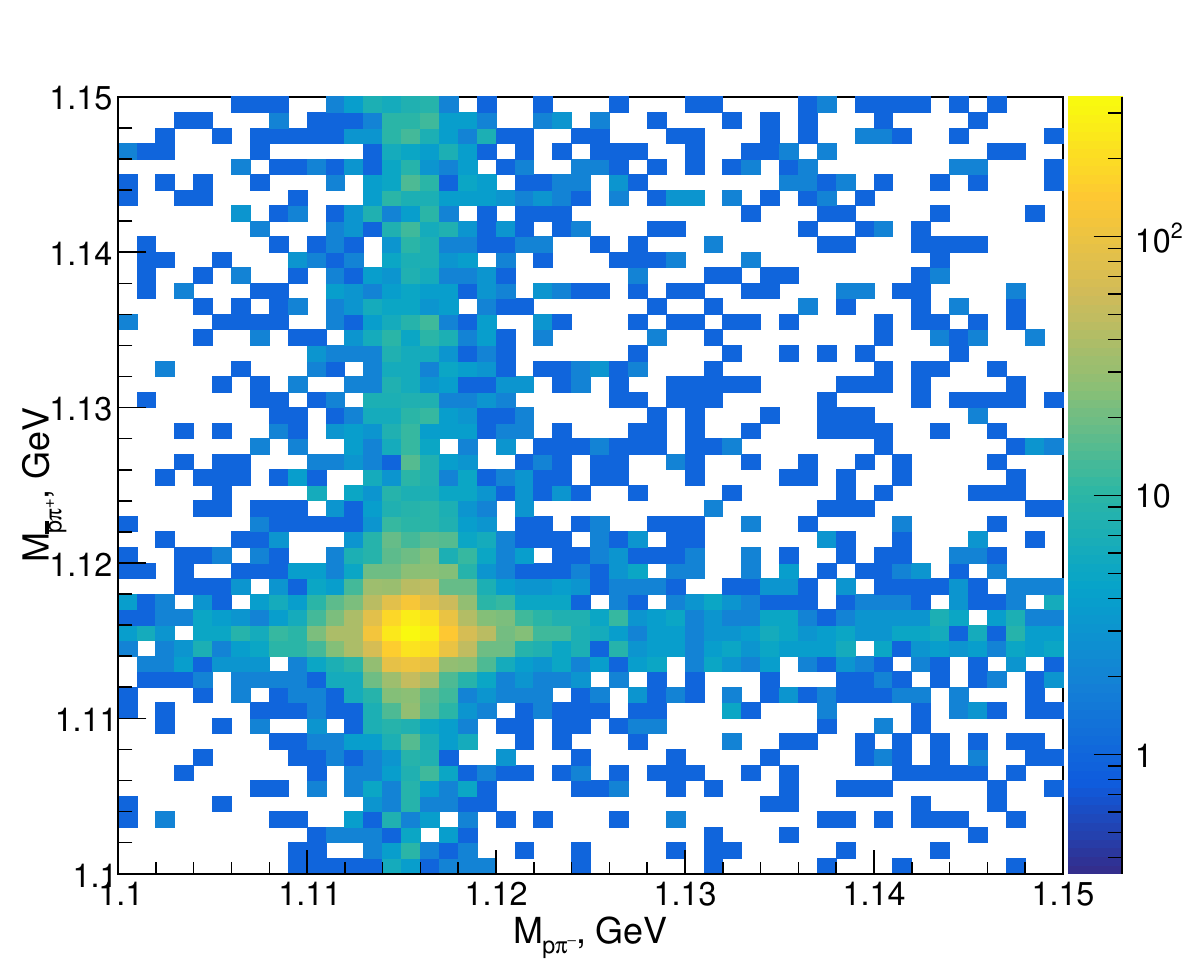} \\ (a)}
  \end{minipage}
    \hfill
   \begin{minipage}[ht]{0.55\linewidth}
   \center{\includegraphics[width=1\textwidth]{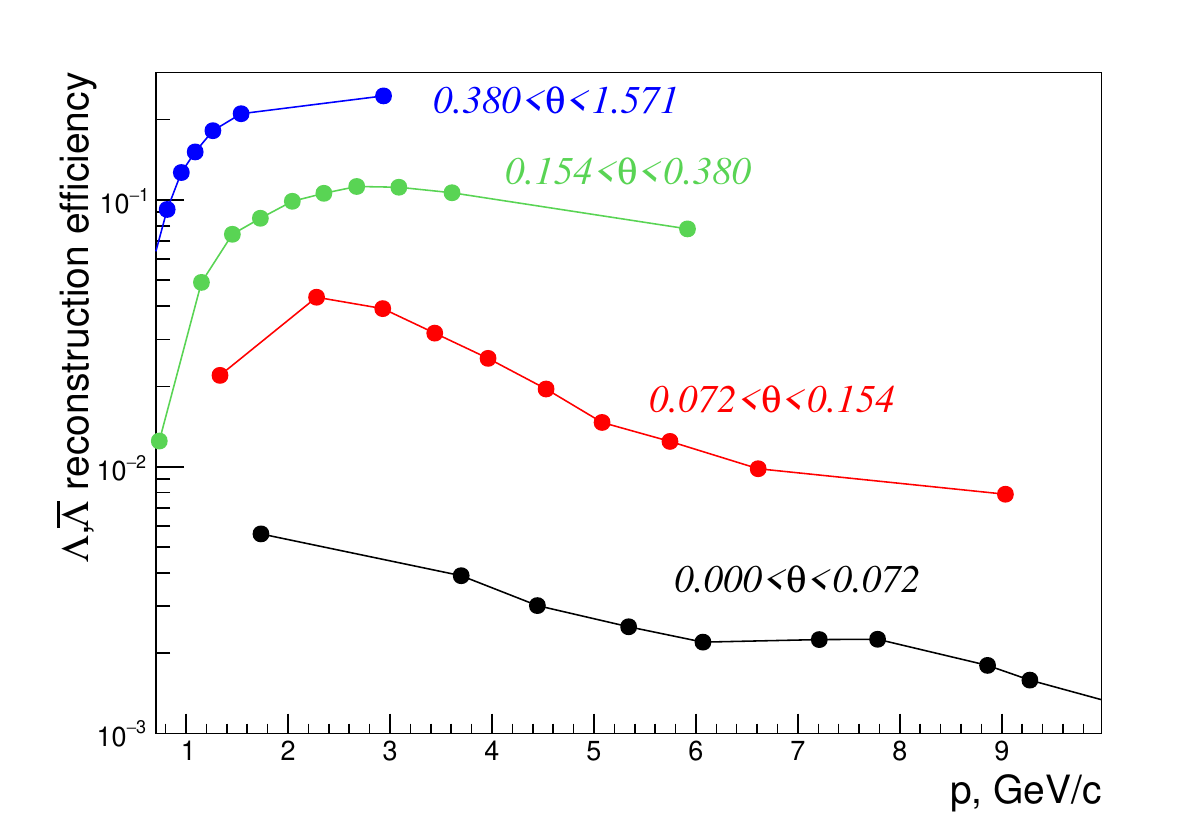} \\ (b)}
  \end{minipage}
  \caption{Distribution of the invariant masses of $\Lambda$ and $\bar{\Lambda}$ candidates in $pp$ collisions at 27~GeV (a). $\Lambda (\bar{\Lambda})$ reconstruction efficiency as a function of momentum and emitting angle with respect to the beam axis (b).}
  \label{lambda_eff}
\end{figure}
The two-dimensional invariant-mass distribution of $\Lambda$ and $\bar{\Lambda}$ candidates 
for $pp$ collisions at 27~GeV is shown in Fig.~\ref{lambda_eff}(a). In the signal-peak 
region, $\pm4$~MeV, the fraction of $\Lambda/\bar{\Lambda}$ signal events reaches 
0.8. The experimental mass resolution is about 1.6~MeV. The reconstruction efficiency of a single $\Lambda (\bar{\Lambda})$ estimated using a full simulation of the SPD detector in its 
complete configuration is presented in Fig.~\ref{lambda_eff}(b)  as a function of momentum and hyperon emitting angle.
The efficiency for the simultaneous reconstruction of the decays 
$\Lambda \to p\pi^-$ and $\bar{\Lambda} \to \bar{p}\pi^+$ was calculated from simulation 
 at a collision energy of 27~GeV and in the initial configuration at 10 GeV. The main event selection 
criteria were the poor association of charged daughter tracks with 
the primary vertex, the distance from the candidate 
$\Lambda (\bar{\Lambda})$ decay vertex to the primary vertex, and 
the angle between the hyperon momentum direction and the vector 
connecting the primary and secondary vertices. Over the full 
kinematic range, the reconstruction efficiency is found to be 
1.5\% and 0.4\% for 27 GeV and 10 GeV, respectively.
In the kinematic region $p_T>0.5$~GeV/$c$ and $|y|<1$, chosen to be similar to that used by STAR, the corresponding values are 5.0\% and 1.8\%. More detailed information for the estimation of the expected statistics is presented in Table~\ref{tab3}.  These calculations do not include the potential inefficiency associated with the software trigger (online filter). The corresponding suppression may reach tens of percent at low energy and up to a factor of several at high energy.

\begin{table}[htp]
\caption{Expected number of reconstructed $\Lambda\bar{\Lambda}$ pairs per month of data taking at 10 and 27~GeV.}
\begin{center}
\begin{tabular}{|l|c|c|c|c|}
\hline
           &  27 GeV & 10 GeV &  27 GeV & 10 GeV \\
\hline
Collisions & $pp$ & $dd$ & $pp$ & $dd$ \\
\hline
Range & \multicolumn{2}{c|}{full} & \multicolumn{2}{c|}{$p_T>0.5$~GeV/$c$ and $|y|<1$} \\
\hline
$\sigma$, $\mu$b & 730&	1400 &	51 &	78 \\
$L$, 10$^{32}$ cm$^{-2}$ s$^{-1}$ &  1 & 0.05 & 1 & 0.05 \\
Efficiency, \% & 1.5 & 0.4 & 5.0 & 1.8 \\
\hline
Offline filter efficiency, \% & \multicolumn{4}{c|}{100} \\
Duty factor, \% & \multicolumn{4}{c|}{70} \\
\hline
Statistics, $10^6$ events & 820 & 30 & 190 & 7.5 \\
\hline
\end{tabular}
\end{center}
\label{tab3}
\end{table}%

 \begin{figure}[!h]
     \begin{minipage}[ht]{0.5\linewidth}
       \center{\includegraphics[width=1\textwidth]{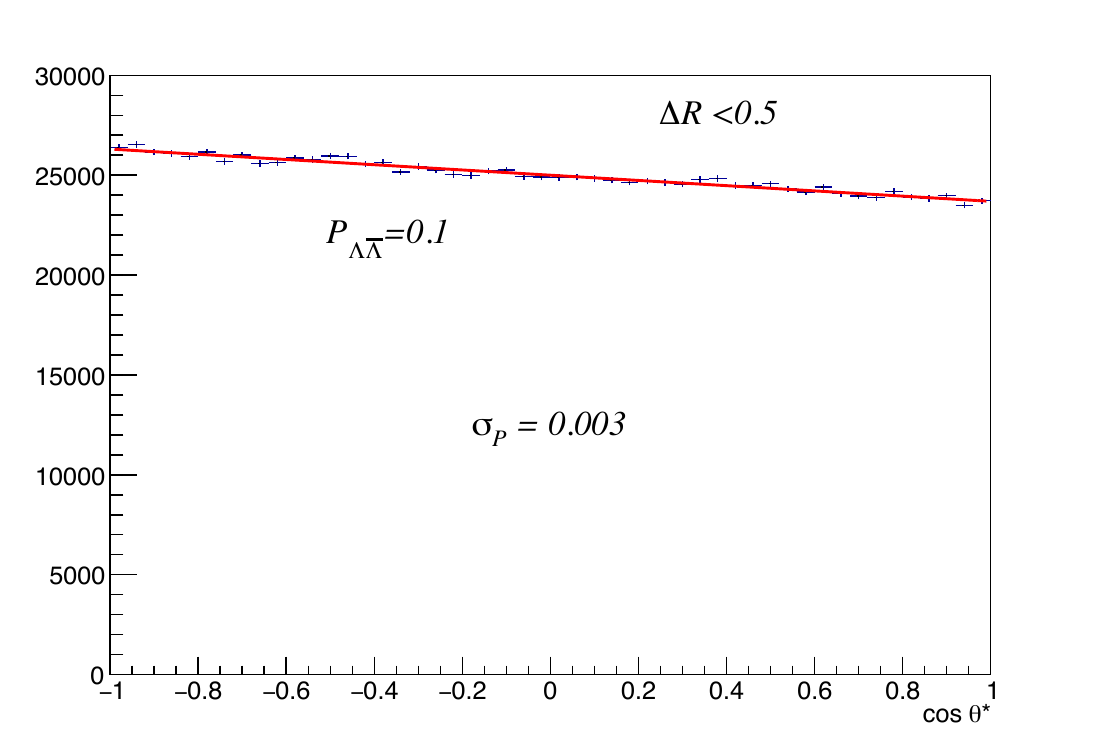} \\ (a)}
  \end{minipage}
    \hfill
   \begin{minipage}[ht]{0.5\linewidth}
   \center{\includegraphics[width=1\textwidth]{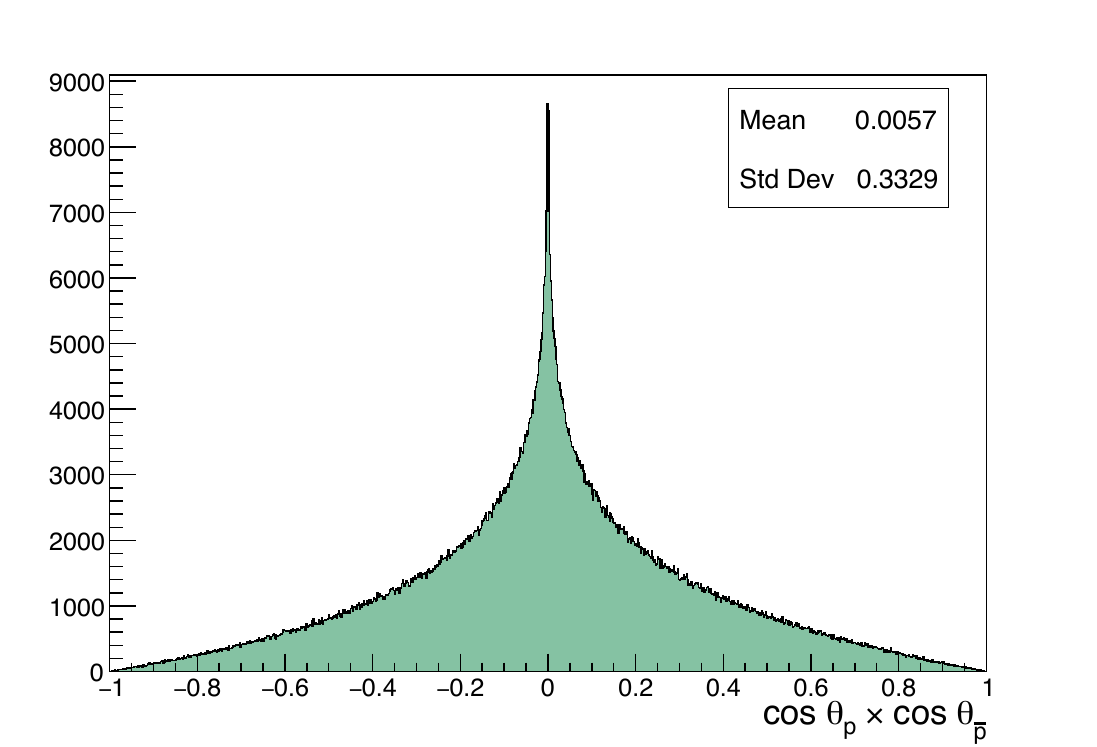} \\ (b)}
  \end{minipage}
  \caption{Expected distributions of $\cos \theta^*$ (a) and $\cos \theta_p \times \cos \theta_{\bar{p}}$ (b) in the kinematic region $p_T>0.5$~GeV/$c$,  $|y|<1$, and $\Delta R<0.5$ for $P_{\Lambda\bar{\Lambda}}=0.1$ and $C_L=0.1$ after one month of data taking in $dd$ collisions at 10~GeV. }
  \label{month}
\end{figure}

Figure~\ref{month} shows how the distributions for $\cos \theta^*$ and $\cos \theta_p \times \cos \theta_{\bar{p}}$ in the kinematic region $p_T>0.5$~GeV/$c$,  $|y|<1$, and $\Delta R<0.5$ with a statistics of $1.2\times10^6$ reconstructed $\Lambda\bar{\Lambda}$ pairs after one month of data taking in $dd$ collisions at $\sqrt{s}=10$~GeV could look. Here  $P_{\Lambda\bar{\Lambda}}=0.1$ and $C_L=0.1$ are assumed. The statistical accuracy of $P_{\Lambda\bar{\Lambda}}$ and $C_L$ extraction is 0.004 and 0.006, respectively.

The main sources of systematic uncertainties in the measurement of 
spin-spin correlations at the SPD experiment are expected to be 
the residual background contributions under the signal peak, 
despite the sideband subtraction, as well as the systematics 
related to the normalisation of the angular distributions. 
Following the approach adopted by STAR, the systematic effects 
can be estimated by analysing the control sample of $K^0_s K^0_s$ 
pairs, where, due to the scalar nature of these mesons, 
no spin correlations are expected to be observed.

The detector acceptance and reconstruction effects, as shown by STAR, can be studied using mixed-event $\Lambda\bar{\Lambda}$ combinations, constructed by pairing hyperons from different events with similar global event characteristics. Such combinations do not contain genuine pair correlations and can therefore be used as a reference for the acceptance-related normalization of the angular distributions.

\section{Conclusion}

In this work, we have studied the feasibility of measuring spin-spin
correlations in $\Lambda\bar{\Lambda}$ pairs at the SPD experiment,
focusing on the intermediate-energy regime accessible at the NICA
collider. The expected production rates, the relative contributions
of different sources of $\Lambda\bar{\Lambda}$ pairs, and the
reconstruction efficiency were evaluated using the \textsc{Pythia} 8
event generator and a full simulation of the SPD detector.

The maximum expected spin-spin correlation
$P_{\Lambda\bar{\Lambda}}$ was estimated within the SU(6) and
Burkardt--Jaffe models for a spin-triplet $s\bar{s}$ state, taking into
account both direct production and feed-down contributions. The
results show that the predicted correlation increases towards lower
collision energies, driven by the rapid decrease of the feed-down
contribution. The $\Xi$ feed-down contribution significantly modifies
the prediction of the Burkardt--Jaffe model because of the opposite
sign of the corresponding spin-transfer coefficient, whereas its effect
on the SU(6) prediction is small. The comparison of the
$\Xi^-/\Lambda$ and $\Sigma^0/\Lambda$ yield ratios with available
experimental measurements provides a consistency check of the
\textsc{Pythia}-based description of strange-hadron production in the
energy range relevant for the study.

The SPD energy range also provides several complementary opportunities
for studying the physics of spin correlations in hyperon production.
Measurements of $P_{\Lambda\bar{\Lambda}}$ as a function of the angular
separation $\Delta R$ at collision energies below those explored by
STAR and CMS can provide a new test of the mechanisms responsible for
the loss of spin correlations during hadronization. In combination
with measurements as a function of event multiplicity, such studies
can help clarify the interplay between the angular separation of the
hyperons and the surrounding hadronic environment in the decoherence
of spin correlations.

The relatively low multiplicities expected at NICA also provide a complementary regime for searching for quantum entanglement in the $\Lambda\bar{\Lambda}$ system. In addition to the spin-spin correlation parameter $P_{\Lambda\bar{\Lambda}}$, measuring the longitudinal correlation $C_L$ would provide access to the spin density matrix under the assumed transverse rotational invariance. The combined measurement of $P_{\Lambda\bar{\Lambda}}$ and $C_L$ would allow the PPT criterion for entanglement to be tested and, if entanglement is present, its magnitude to be quantified through the negativity. This provides a stronger test of the quantum nature of the $\Lambda\bar{\Lambda}$ correlations than $P_{\Lambda\bar{\Lambda}}$ alone.

Furthermore, the dependence of the correlations on collision energy,
kinematics, and beam polarization can provide sensitivity to different
$s\bar{s}$ production mechanisms, including non-perturbative production,
gluon splitting, and hard scattering. In particular, measurements with
longitudinally and transversely polarized beams can probe the role of
initial-state polarization and spin-dependent partonic dynamics.
Measurements with different $\Lambda$ parentage can additionally
constrain the transfer of spin information from the $s$ quark to the
final-state hyperon and help distinguish direct production from
feed-down contributions.

These results demonstrate that the SPD experiment can extend the study
of hyperon spin correlations to a previously unexplored energy domain.
The combination of collision-energy, angular-separation, multiplicity,
kinematic, and beam-polarization dependences makes SPD a promising
facility for studying the evolution of spin correlations from
$s\bar{s}$ production to the final-state $\Lambda\bar{\Lambda}$ system,
including their possible quantum origin, decoherence during
hadronization, production mechanisms, and spin transfer.

\end{document}